\documentclass[12pt]{iopart}   
\usepackage{epsfig}          

\def\d{\rm d}
\def\aunits{\rm cm^{-2}s^{-1}sr^{-1}GeV^{-1}}
\newcommand{\gsim}{\lower.7ex\hbox{$\;\stackrel{\textstyle>}{\sim}\;$}}

\begin{document}

\title{Measuring Diffuse Neutrino Fluxes with IceCube}
\author{Marek Kowalski}
\address{ Lawrence Berkeley National Laboratory 
\\ Berkeley, CA, 94720, USA \\
Email:  {\tt MPKowalski@lbl.gov}}

\begin{abstract}

In this paper the sensitivity of a 
future kilometer-sized neutrino detector to detect and 
measure the diffuse flux of high energy neutrinos is evaluated. Event rates 
in established detection channels, such as muon events from charged current 
$\nu_\mu$ interactions or cascade events from $\nu_e$ and  $\nu_\tau$ interaction, are 
calculated using a 
detailed Monte Carlo simulation. Neutrino fluxes as 
expected from prompt charm decay in the atmosphere or from astrophysical 
sources 
such as Active Galactic Nuclei (AGN)  are modeled assuming power laws.
The ability to measure the normalization and slope of these spectra is then 
analyzed. 

It is found that the cascade channel generally has a high sensitivity for the 
detection and characterization of the diffuse flux, when compared 
to what is expected for the upgoing- and downgoing-muon channels. 
A flux at the level of the Waxman-Bahcall upper bound should be detectable 
in all  
channels separately while a combination of the  information of the different 
channels will allow detection of a flux more than one order of magnitude 
lower. 
 Neutrinos from the prompt decay of  charmed mesons in the 
atmosphere should be detectable in future measurements for all but the lowest 
predictions.   

\end{abstract}

\maketitle

\section{Introduction}
High energy extra-terrestrial neutrinos have so far escaped their detection, 
and there is a considerable effort invested to change this situation soon 
\cite{spier}. 
IceCube, a Cherenkov detector with a volume of a cubic kilometer is currently 
being installed at the South Pole \cite{henrike} and plans for a northern 
hemisphere 
detector of similar size are maturing \cite{antares}. Such detectors are 
expected to measure 
high energy neutrinos ($E_\nu$ \gsim 100~GeV) from various sources of which
only a few might be resolved by directional information. The unresolved 
{\it diffuse} (or isotropic) neutrino flux is of vital interest as well as 
it could reveal sources typically associated with very distant and energetic 
astrophysical objects such as AGNs, GRBs and 
even first signs for new physics beyond the Standard Model 
(see \cite{lm,hh} for a review of current models). 
 
However, the diffuse flux must be observed above the background of 
atmospheric neutrinos. The so called conventional atmospheric neutrino flux  from pion and kaon decay produced in cosmic ray interactions in the atmosphere falls steeply with energy ($\frac{\d N}{\d E_\nu}\propto E_\nu^{-\gamma},\gamma\approx3.7$ for $E_\nu \gsim 10$~ TeV). At higher energies 
$\mathcal{O}(10^5 \mbox{ ~GeV})$ the contribution of neutrinos
 from the prompt decay of charmed mesons will begin to dominate the flux, 
thereby hardening its spectral index 
to $\gamma \approx$2.7-3.0. The neutrino
contribution from charm decay is yet unobserved and predictions vary by more 
than one order of magnitude. A measurement would allow to infer
the cross-section for production of charmed particles.

Hence there are a number of interesting potential contributions to the diffuse flux of high energy neutrinos. In 
this paper we address the question under which conditions the different contributions can be resolved in future measurements. A 
Likelihood function is constructed incorporating both observable 
energy and angular information as well as  
systematic uncertainties of the background.  The information content of 
future measurements is then analyzed by means of the Fisher matrix technique. 
We confirm the well know fact, that a cubic kilometer sized detector such as 
IceCube has a large discovery potential. 
However, at the same time we show that it also has a significant 
discrimination potential with respect to  the various models. We analyze the information 
content contained in the various detection channels, namely the  
cascade- and muon-channels, and show what can be 
gained from their combination.

This paper is organized as follows. In the next section the various 
fluxes, experimental detection channels as well as the simulation of 
event rates are being discussed. In section \ref{sec:fisher} we review the
Fisher matrix method and present its implementation with respect to the 
problem at hand. In section \ref{sec:results} we show the results of this 
analysis for a number of different cases. 
We conclude with a discussion of the results and compare to previous work done 
on similar subjects.

\section{Simulation}
\label{sec:simulation}
The following fluxes will be considered in various combinations 
throughout the paper. 
\begin{itemize}
\item An astrophysical flux of neutrinos: 
$\d N/\d E = \alpha_{\rm agn}\times E_\nu^{-\gamma_{\rm agn}}$. The slope $\gamma_{\rm agn}$ as well as the 
normalization $\alpha_{\rm agn}$ are treated as 
unknown parameters. The assumed true value for the slope is 
$\gamma_{\rm agn}=2$ while 
$\alpha_{\rm agn}$ 
is allowed to vary from $10^{-9}$ to $10^{-7}~\aunits$.
The upper limit is chosen slightly below present 
experimental upper bounds \cite{marek,gary,hundert,baikal}. The Waxman-Bahcall (WB) upper 
bound \cite{wb} (see also \cite{mpr}), which is of significant astrophysical relevance, corresponds to 
$\alpha_{\rm agn}=4\times 10^{-8} \aunits$. 
Neutrino oscillations will result in approximate 
equipartition of the total neutrino flux among all 
flavors \cite{Athar:2000yw}. $\alpha_{\rm agn}$ refers to the 
normalization of the flux of one flavor.

\item A flux of neutrinos from charm decay in the atmosphere. This flux is 
           approximated by a power-law spectrum with $\gamma_{\rm charm}=2.8$. 
The 
normalization $\alpha_{\rm charm}$ ranges from $10^{-5}$ to $10^{-3}~\aunits$, 
reflecting the range of available models around 100 TeV neutrino energies
\cite{costa,showerpower,Candia:2003ay}.
Although the shape of the spectrum is generally rather well predicted, 
small variations arise for example from the use of different cross-section 
parameterizations. Hence we again leave both normalization and 
slope as free  parameters. A similar flux is assumed for 
$\nu_e$ and $\nu_\mu$ with no  contribution from $\nu_\tau$.

\item The atmospheric neutrino flux. Here we use a selected
 model \cite{lipari} and we do not assume any free parameters. 
It was shown that the uncertainty in the flux prediction degrades 
the ability to detect additional flux contributions \cite{hooper}.
A 15\% 
systematic uncertainty in the normalization is assumed \cite{honda}. The 
normalization is allowed to 
vary as a function of energy with a correlation length of one decade in 
energy. The assumed correlation in energy of the uncertainty 
reflects the fact that relevant ingredients of the calculations, as for 
example the 
cosmic ray composition or the interaction cross-sections of mesons in the 
atmosphere, are not anticipated to show sudden 
unaccounted variations. Hence in an experiment, the systematic 
uncertainties between two neighboring energy bins is correlated. 
The implementation of this correlated uncertainty is discussed in 
section \ref{sec:fisher}.

Note that neutrino oscillations have a very small effect on the 
atmospheric neutrino flux, due to the high energies considered here. 
\item The atmospheric muon flux. An analytical parameterization was used
 \cite{pdg}, which even at energies of $10^7$~GeV agrees within a 
factor of two with a full  CORSIKA Monte Carlo simulation \cite{shigeru}. 
The flux of atmospheric muons at the energies of interest ($>$PeV) is less well 
predicted, because of the high energies involved. We assume an uncertainty of 
50~\%, again with a correlation length of one decade in energy.
\end{itemize}

Neutrino events ranging from $10^3$ to $10^{11}$~GeV are simulated using the 
Monte-Carlo simulation program ANIS \cite{anis}.
ANIS allows generating $\nu$-events of all
flavors, propagates them through the Earth and finally simulates 
$\nu$-interactions within a specified volume. All
relevant Standard Model processes, i.e.\ charged current (CC) and neutral current (NC) 
$\nu N$-interactions as well as resonant $\bar \nu_e-e^-$ scattering are implemented. Neutrino
regeneration as expected in NC-scattering, $\nu + N \rightarrow \nu +
X$, and in $\tau$ production and decay chains, $\nu_\tau + N
\rightarrow \tau + X$,  $\tau \rightarrow \nu_\tau + (\nu_i) + X$, are
included at all orders. The density profile of the Earth is chosen according 
to the Preliminary Earth Model \cite{prem}. 
Deep inelastic $\nu-N$-cross-sections were calculated in the framework of 
perturbative QCD (pQCD).
Parameterization of the structure functions were chosen 
according to CTEQ5 \cite{cteq}, with logarithmic extrapolations into the 
small-$x$ region. Tau decay was simulated using the TAUOLA program \cite{tauola}.

The muon energy  after propagating a distance $X$ is approximated
 by: $E_\mu=(E_{\mu,0}-a/b)e^{-bX}-a/b,$
with $a=3$~MeVg$^{-1}$cm$^{2}$ and $b=4\times10^{-6}$~g$^{-1}$cm$^{2}$ \cite{pdg}. Note 
that we are neglecting the fluctuations in muon range leading to slightly 
over-optimistic sensitivities for muons. 

An IceCube-like detector is simulated. It is embedded in ice 
at a depth of 2~km with a  bedrock starting  in 3~km depth. 
We define three classes of 
events, all of which have been used by the AMANDA collaboration
 to search for a diffuse flux of neutrinos: cascades, upgoing muons and 
downgoing muon events. These classes are explained below. 

\begin{itemize}
\item {\it Cascade} events which consist of electro-magnetic or hadronic cascades (also named showers). The particle cascade is localized in space and hence can only be observed near or within the detector volume. Charged current $\nu_e$ and $\nu_\tau$ interactions as well as NC interactions of any neutrino flavor lead to cascade-like events. In case of NC interactions, the visible energy is 
just a fraction of the incoming neutrino energy. In case of a CC tau neutrino interaction  the vertex cascades and the cascade obtained from tau decay are 
spatially separated  \cite{doublebang}. It is not yet established under which 
conditions and at which energy this {\it double bang} signature 
can be experimentally resolved, hence tau neutrino identification will not 
be assumed.

A main advantage of the cascade channel is the reduced background of 
atmospheric muons.
An analysis performed with AMANDA has shown that for this class of events an 
almost uniform directional 
sensitivity can be reached  \cite{marek,marek_phd}. This  leads to an improved sensitivity at higher energies, where Earth absorption effects become relevant. It is assumed that the background of atmospheric muons, due to their different 
event topology, can be totally eliminated above 1~TeV cascade energies.

\item {\it Upgoing muon} events from CC $\nu_\mu$ interactions producing an 
energetic muon transversing the 
detector. These events are selected by a zenith angle cut, 
$\theta > 90^\circ$, to eliminate the background of atmospheric muons.
This is the traditional observation mode. The vertex of such muon events 
can be many kilometers away, hence the event rate for this class of 
events is generally higher than in the case of cascades. A further advantage 
is that for these events an 
angular resolution of about 1 degree can be achieved \cite{henrike}.
At high energies, neutrino absorption reduces the event rates in this channel 
significantly.
\item{\it Downgoing muon} events from CC $\nu_\mu$ interactions 
do not suffer of absorption in the Earth, hence they allow to extend the 
sensitivity of the instrument to {\it ultra high} energies beyond $10^9$~GeV 
\cite{hundert,shigeru}. The disadvantage of this channel is that there is a 
large  background of atmospheric muons. 
Therefore the effective energy threshold of this channel is $10^6-10^7$~GeV.
Here we will assume that the direction of these events can be reconstructed 
with a resolution of better than $5^\circ$. This is not yet achieved for 
AMANDA, however the large increase in size and  modern readout technology should easily allow IceCube to achieve this resolution.
\end{itemize}

\begin{figure}[t]

  \centering
    \includegraphics[width=\linewidth]{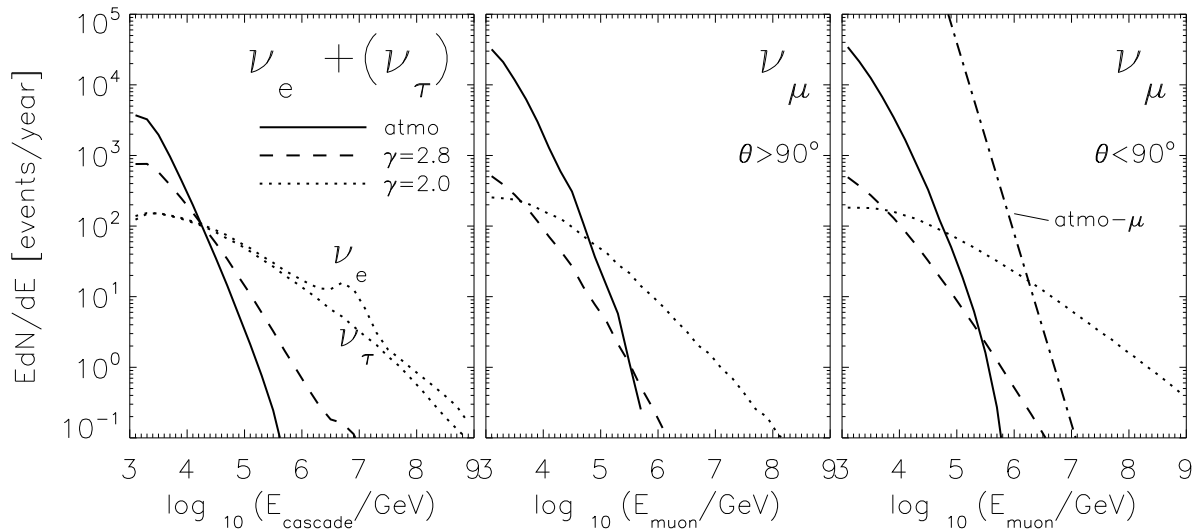}%
  \caption{Observable energy spectrum at the detector site. An energy 
resolution of $\sigma_{\log E}=0.2$ was assumed. 
Left:  The cascade energy released 
by electron neutrino interaction within a 1~km$^3$ sized detector for three 
different input spectra and averaged over all directions.
Also shown is the contribution expected from tau neutrinos with a flux 
similar to that of $\gamma=2$ electron neutrinos. 
Middle: 
The upgoing 
muon event rate as a function of the muon energy at the detector. 
A detection area of 
1~km$^2$ was assumed. Right: The downgoing 
muon event rate as a function of the muon energy at the detector. Additionally shown is the event rate due to atmospheric muons.
The two input power-law spectra  have slopes of $\gamma=2.0 (2.8)$ and
 normalizations $\alpha = 4\times 10^{-8} (5\times 10^{-4}) \aunits$.}
    \label{fig:spec}

\end{figure}

The resulting event rates are shown as a function of the observable energy 
in figure~\ref{fig:spec}. The left plot shows event rates for electron 
neutrinos for different input spectra. Also shown are 
event rates for tau neutrinos 
from an astrophysical flux with $\gamma_{\rm agn}=2.0$.  The ``bump'' 
around $6\times10^6$~GeV is due to resonant $\bar\nu_e-e^- $ scattering. 
For the highest energies the rate of tau neutrinos drops below that of electron neutrinos. That 
is because  of the increasing tau decay length. The tau decays more 
frequently outside (``behind'') the 
detector, hence it does not contribute to the visible energy. The 
middle plot show the event rates for upgoing neutrino-induced muons as a 
function of the muon energy in the detector. The 
right plot show the event rates for downgoing neutrino-induced muons as well 
as for atmospheric muons.

An energy resolution of $\sigma_{\log E} \approx 0.2 $ for all event classes is assumed. Current algorithms  
used for the operating AMANDA detector are already achieving this resolution for cascade-like events 
 \cite{marek,marek_phd}.  For muon events this is a mild extrapolation of the 
currently obtained resolution: $\log E_\mu \approx 0.3$  \cite{heiko,ped}. However, the results of this work are not very sensitive on the exact value of the energy resolution.

\section{The Method}
\label{sec:fisher}
One of the most popular and powerful methods for parameter estimation is the {\it maximum Likelihood method}, in which the parameters are determined by maximizing a Likelihood function $L$ with respect to its parameters $\vec{p}$.
Because of the low neutrino event rates, the  appropriate 
Likelihood function is based on Poisson statistics:
\begin{equation}
L(\vec{p})=\prod_{i} P(N_i|n_i(\vec{p})) = \prod_{i} 
\frac{N_i^{n_i} e^{-n_i}}{N_i !},
\label{eq:like} 
\end{equation}     
where $i$ is the index of the data bin, $N_i$ is the number of 
observed events in data bin $i$ and $n_i$ is the number of events expected given the parameters $\vec{p}$.
It is convenient to define the Log-Likelihood function as $\mathcal{L}\equiv-\log L$ so that for the Likelihood function defined in Equation (\ref{eq:like}) one obtains: $\mathcal{L}=-\sum_iN_i\ln n_i - n_i - \ln N_i!$.

Here we are interested in the achievable accuracy of possible future 
measurements. To evaluate the sensitivity of the upcoming instruments 
one assumes a fiducial model, represented by  
$\vec{p}$, and analyzes the expected averaged constrains on the input parameters. However,
  there 
is more than one way to do this. One can simulate a future experiment in  
 detail and then average over a large number of experimental 
outcomes by means of Monte Carlo technique. 
Such an approach produces very accurate results but at the same time it is 
computationally 
demanding and cumbersome in its  implementation. 

Here we chose a different method to calculate the sensitivity, 
namely the Fisher matrix method \cite{fisher}. This method has become a
popular tool in modern cosmology (see for example \cite{tegemark}), most 
likely due to its transparency and computational efficiency. 
Curiously, it is rarely used in the field of 
astroparticle physics. Hence we briefly review this method here (see 
\cite{tegemark} for a more in depth discussion of the method). 

The Fisher information matrix is defined as the ensemble averaged Hessian 
matrix of the Log-Likelihood function evaluated at its minimum, 
\begin{equation}
F_{\rm ij} \equiv \langle \frac{\partial^2 \mathcal{L}}{\partial p_{\rm i} \partial p_ {\rm j}} \rangle. 
\end{equation}

It represents the information content of the Likelihood function
in the close vicinity of the true parameters. There are two 
important inequalities: $\sigma_{p_i} \le 1/\sqrt{F_{ii}}$, a statement 
about the lower limit obtainable on the uncertainty $\sigma_{p_i}$ on the 
parameter 
$p_i$ in the case that all other parameters are exactly known. This theorem is 
known as the Cram$\acute{\rm e}$r-Rao inequality. The second inequality is: 
$\sigma_{p_i} \le \sqrt{(F^{-1})_{ii}}$, which is applicable in the case of all
parameters being unknown (the parameters $j\neq i$ are 
marginalized over). Here we will assume that $\sigma_{p_i}=
\sqrt{(F^{-1})_{ii}}$, which
 is a reasonable approximation for well behaved Likelihood-functions.

For the Poisson based Likelihood function (\ref{eq:like}), one 
obtains the simple expression: 
\begin{equation}
F_{ij} = \sum_{ k} \frac{1}{n_{ k}}\frac{{\partial} n_{ k}}{{\partial} p_{ i}}\frac{{\partial} n_{k}}{\partial p_{ j}}
\end{equation}

Note that so far only statistical errors have been assumed. What follows is
 an attempt to incorporate anticipated systematic uncertainties in the 
background prediction. We assume that the conventional atmospheric neutrino 
flux normalization at all energies is known with a precision of 
$\sigma_{\rm atmo}=15\%$ \cite{honda}, and that this 
uncertainty has a correlation length of one decade in energy. 
The uncertainty alters the neutrino event rate prediction $n_i^\prime=(1+\eta_i) n_i$, where $\eta_i$ is a nuisance parameter reflecting the systematic uncertainty. Its probability function, which needs to be multiplied with the right hand side of Equation (\ref{eq:like}), is assumed to be Gaussian:
\begin{equation} 
P(\vec{\eta})
\propto e^{-\vec{\eta}C\vec{\eta}} \mbox{ with }
C_{ij}=\frac{1}{2\sigma_{\rm atmo}^2(1+|\log (E_i/E_j)|)^2}.
\end{equation}

The dimension of the Fisher matrix increases by the number of nuisance parameters. Assigning the indices $i,j$ to the physical 
parameters of interest (e.g. flux normalization or spectral slope) and  $\rho,\eta$ to the nuisance parameters, one obtains the 
following additional Fisher matrix elements:
\begin{equation}
F_{\rho\eta} = \sum_k C_{\rho\eta}+\delta_{\rho\eta} \frac{(n^{\rm atmo}_k)^2}{n_k}
\mbox{ and } F_{i\delta}=  \sum_k \frac{n^{\rm atmo}_k}{n_k}\frac{\partial n_k}{\partial p_i}. 
\end{equation}

\section{The sensitivity}
\label{sec:results}
As described in section \ref{sec:simulation}, we 
assume a neutrino flux composed of three components: 
\begin{equation}\phi(E_\nu)= 
\phi(E_\nu)_{\rm atmo}+\alpha_{\rm agn} E^{-\gamma_{\rm agn}} +\alpha_{\rm charm} E^{-\gamma_{\rm charm}},\end{equation} with $\gamma_{\rm agn},\gamma_{\rm charm},\alpha_{\rm agn}$ and $\alpha_{\rm charm}$ being the free parameters of 
interest. Additionally there is a flux of atmospheric muons, which is the relevant background for the downgoing muon channel.

For this analysis the data was binned in energy-bins of the size of 
the energy resolution: $\Delta\log E=0.2$. It was found that for the cascade and upgoing muons  channel, there is little 
sensitivity increase when binning the data in energy {\it and} zenith 
angle direction. However, in the case of downgoing muons, the sensitivity generally increases by about a factor of two, if the reconstructed zenith angle is used as additional information (with bins of $\Delta\cos\theta$=0.1). 
 This is because of the increase in overburden towards horizontal 
directions. The background of atmospheric muons 
decreases fast towards the horizon, while the signal rate increases due to 
increased target material. Hence for the downgoing muon channel explored in 
this 
work, the additional directional information is being used.

\begin{figure}[t]
  \centering
    \includegraphics[width=\linewidth]{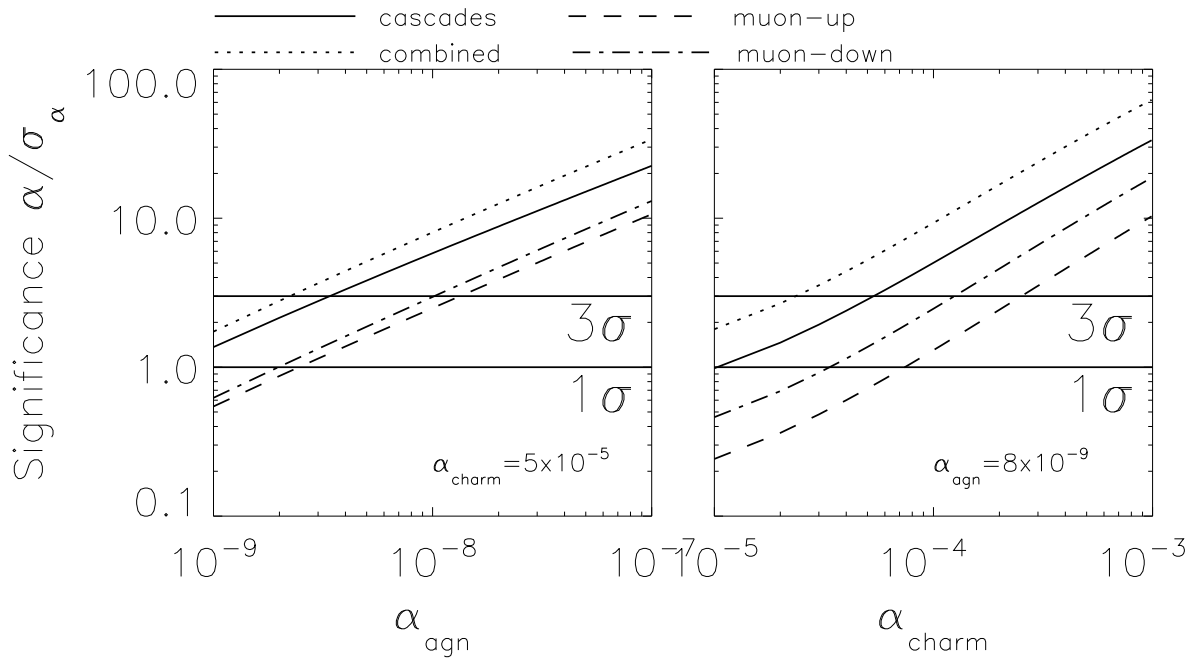}
  \caption{Significance (defined as $\alpha/\sigma_\alpha$) of the measurement of $\alpha$ as a function of $\alpha$. Left: The significance for $\alpha_{\rm agn}$ assuming an additional contribution of neutrinos from charm decay 
at the level of $\alpha_{\rm charm} = 5\times 10^{-5} \aunits$. Right: The significance of $\alpha_{\rm charm}$ assuming an extraterrestrial
 contribution with slope $\gamma_{\rm agn}=2$ and normalization 
$\alpha_{\rm agn} =8 \times 10^{-9} \aunits$.  }
\label{fig:alpha}
\end{figure}

Figure \ref{fig:alpha} shows the significance 
$\alpha/\sigma_\alpha = \alpha \sqrt{(F^{-1})_{\alpha\alpha}}$ 
 with which the normalization $\alpha$ can be measured as a function of 
$\alpha$ assuming three full years of data. The various 
lines correspond to the different detection channels.
The dotted line represents the combined measurement, where the corresponding 
Fisher matrix is\footnote{Combining the individual Fisher matrices in a simple sum  is only valid if there are no correlations between the channels. Certainly there will be some common systematic uncertainties, so the simple sum should be considered an optimistic assumption.}
$F_{\rm combined}=F_{\rm cascade}+F_{\rm up-muon}+F_{\rm down-muon}$.

\begin{figure}[t]
  \centering
    \includegraphics{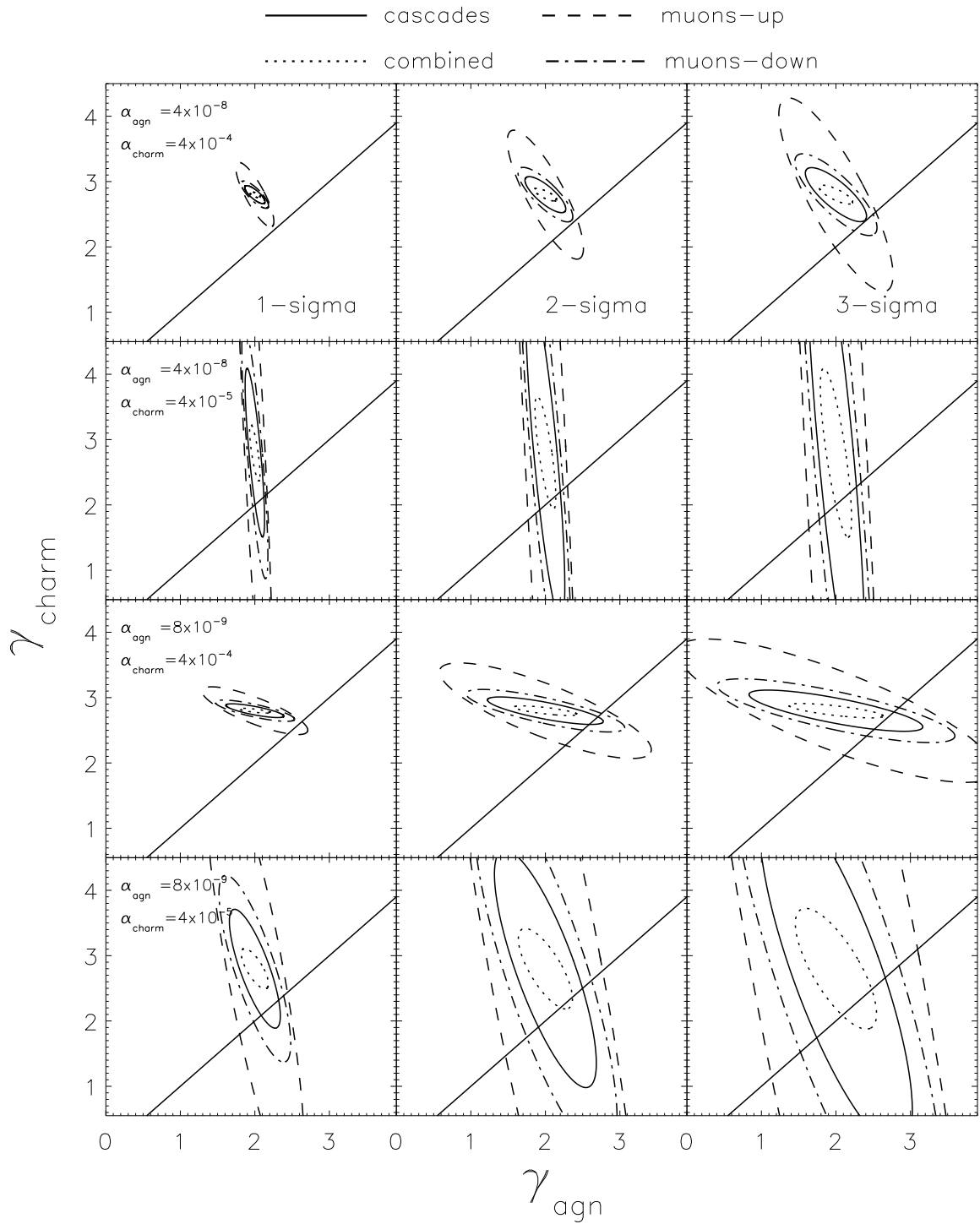}%
  \caption{Averaged error ellipse in the $\gamma_{\rm agn}-\gamma_{\rm charm}$
 plane for the different channels. The columns from left to right show 1-,2- and 3-sigma contours. The different rows show various combinations of flux normalizations.
The diagonal lines represent $\gamma_{\rm agn}=\gamma_{\rm charm}$.}
\label{fig:gamma}
\end{figure}

The left figure shows the significance of the extraterrestrial component 
with a charm contribution of strength $\alpha_{\rm charm}=5\times 10^{-5}~\aunits$ 
while the
 right figure shows the significance of the measurement of the 
charm component with an assumed 
extraterrestrial contribution of strength 
$\alpha_{\rm agn}=8\times10^{-9}~\aunits$.

These curves have a simple interpretation. We assume a measurement is called a 
discovery once the signal is established with 3-$\sigma$ significance (or in other words incompatible with zero at the 99.73~\% level).  
One can read off the 
corresponding $\alpha$ for which an ensemble of experiments will report a 
discovery. For smaller values of $\alpha$ one will 
 typically obtain only upper limits, while for larger values of $\alpha$ a 
discovery becomes more significant. 

As can be seen, a flux at the WB bound would be clearly detected in all three 
channels. The cascade channel, being most sensitive,  will allow to make a 
detection at the 
$\alpha=6\times10^{-9}$ level, while the combined sensitivity still
improves the results notably.

A flux of neutrinos from charm decay will be detected in all 
channels only for the highest of all normalizations. Here 
the cascade channel is significantly more sensitive allowing to detect
a flux for $\alpha_{\rm charm}$ as low as $5\times10^{-5}~\aunits$. However, in the presence of a 
significant  flux of astrophysical neutrinos with spectral index 
$\gamma\approx 2$, 
it will become difficult to probe the lowest of all model predictions.

Note that a discussion on the expected uncertainties 
on the normalization itself becomes more complicated because of 
the strong covariance between the spectral shape and the normalization 
$\alpha$.   The covariance is large, since $\alpha$ is defined for 1~GeV while
the relevant energy scale\footnote{A representative energy scale can
be defined as the weighted average of the energy of the bins, with the weights
given by the bins
contribution to the Fisher matrix.} $E_s$ is 100~TeV to 1~PeV. Redefinition 
of the normalization to $\alpha_s=\alpha E_s^{\gamma}$ would 
reduce the covariance. By presenting the sensitivity 
as $\alpha/\sigma_\alpha$, the results are independent of the energy 
scale assumed.

Once a signal has been established the shape of the underlying spectrum 
becomes of interest. 
Figure \ref{fig:gamma} shows the error ellipses in 
the $\gamma_{\rm agn}-\gamma_{\rm charm}$ plane 
obtained for different combination of 
small/large extraterrestrial/charm fluxes. The normalization values 
used are $\alpha_{\rm agn}=(0.8 - 4)\times10^{-8}~\aunits$ and $\alpha_{\rm charm}=(0.4 - 4)\times10^{-4}~\aunits$. 
The different columns show different 
significance contours ranging from 1- to 3-sigma (from left to right).
The diagonal lines represent $\gamma_{\rm agn}=\gamma_{\rm charm}$. 
The interpretation of these ellipses are -- within the framework of the Fisher matrix method -- the following: Given the true values of $(\alpha_{\rm agn}, \gamma_{\rm agn})$ and $(\alpha_{\rm charm}, \gamma_{\rm charm})$, future measurements of these parameters will be distributed according the probability contours. 

In case  the result of a measurement intersect the diagonal line, 
separation of the two components
becomes impossible. As can be seen in the top row of the the figure, 
if both contributions are large, separation is possible at the 3-sigma level. 
If $\alpha_{\rm agn}$ is large, but $\alpha_{\rm charm}$ is small, only 
$\gamma_{\rm agn}$ can be meaningfully  constrained. This is not surprising, 
considering that $\alpha_{\rm charm}$ has not yet been 
measured with 3-sigma significance. The other two combinations both 
show at most a 2-sigma separability of the two components. Again, of the 
individual channels it  is the cascade channel which provides the tightest 
constrains.

\section{Discussion}
This paper gives an outlook on the sensitivity of future 
measurements of the diffuse neutrino flux. 
The primary detection channels used by AMANDA, which have proven capabilities \cite{marek,hundert,gary}, have been investigated and their sensitivity for 
measuring normalization and spectral shape of various potential fluxes was  
evaluated for a kilometer sized detector.

Thereby it was found that the cascade channel has the highest 
sensitivity. With three years of data, it should be possible to discover a flux 
of extraterrestrial neutrinos at a level as low as 
 $4\times10^{-9}~E^{-2}~ \aunits$. It was found that the downgoing 
muon channel is rather sensitive, too, if 
in addition to muon energy the directional information is used.  The 
traditional way to search for high energy neutrinos, namely by identifying 
upgoing muons, is least sensitive. This is because neutrino absorption in the Earth 
suppresses the observable neutrino spectrum at higher energies, therefore  
reducing significantly the lever arm needed to determine the spectral shape.

A combination of the individual channels leads to the largest sensitivity.
Nevertheless, for confirmation it would be desirably to detect a significant 
signal in at least two of the three individual channels.

A subset of the results presented here has been discussed previously. A detailed  
simulation of the muon channel for IceCube has resulted in precise estimates of the 
sensitivity to a single component diffuse flux \cite{henrike}. However, spectral shape 
determination and the separability of an astrophysical flux contribution from a contribution from 
 atmospheric prompt charm decay were not discussed. It was 
pointed out before that the cascade channel has a significant potential \cite{showerpower,marek_phd,hooper}.
The ability to determine the spectral shape using the cascade channel was 
studied in Hooper 
et {\it al.} \cite{hooper}. Because neutrino propagation in the Earth was 
not included in their 
simulation the analysis was restricted to downgoing cascades only. Further only CC interactions were 
taken into account. Finally the energy bins were chosen to be one order of magnitude, hence significantly 
larger than the energy resolution and bin size assumed here, reducing the sensitivity further. 
This list of differences explains why comparable sensitivities are obtained assuming only three 
years of data here while ten years of data was assumed in \cite{hooper}. 

The work presented here is a first attempt to compare the various detection 
channels in a quantitative manner.
Future analyzes might improve upon the sensitivities 
presented here. 
If tau neutrinos could be identified as 
such it would provide an important additional signature \cite{beacom_tau}, as 
there is  negligible background due to atmospheric tau neutrinos. Hence a 
single event would be a significant discovery.  
In practice the effective volume/area of an analysis is not confined to the 
geometrical one. For energetic -- hence bright -- events it could be larger, 
leading to higher sensitivities.
Finally, one should restate that the analysis presented here assumes the 
availability of three years of data, 
while IceCube operation is planned for a time period of 10 years.

\vspace{0.5cm}

\noindent
{\bf Acknowledgments} 

\noindent
The author would like to thank J.~Beacom, T.~Gaisser, T.~Hausschildt and C.~Spiering for helpful discussions and A.~Gazizov for a fruitful collaboration on the ANIS program.


\vspace{0.5cm}
\begin{thebibliography}{99}
\bibitem{spier}
  C.~Spiering,
  Preprint arXiv:astro-ph/0503122.
\bibitem{henrike}
  J.~Ahrens {\it et al.}  [IceCube Collaboration],
  Astropart.\ Phys.\  {\bf 20}, 507 (2004).

\bibitem{antares}
  P.~Piattelli  [NEMO Collaboration],
  Nucl.\ Phys.\ Proc.\ Suppl.\  {\bf 138} (2005) 191.
\bibitem{lm}
  J.~G.~Learned and K.~Mannheim,
  Ann.\ Rev.\ Nucl.\ Part.\ Sci.\  {\bf 50} (2000) 679.

\bibitem{hh}
  F.~Halzen and D.~Hooper,
  Rept.\ Prog.\ Phys.\  {\bf 65} (2002) 1025.


\bibitem{marek}
  M.~Ackermann {\it et al.} [AMANDA Collaboration],
  Astropart.\ Phys.\  {\bf 22} (2004) 127.

\bibitem{gary}
  J.~Ahrens {\it et al.} [AMANDA Collaboration],
  Phys.\ Rev.\ Lett.\  {\bf 90} (2003) 251101.

\bibitem{hundert}
  M.~Ackermann {\it et al.} [AMANDA Collaboration],
  Astropart.\ Phys.\  {\bf 22} (2005) 339.

\bibitem{baikal}
J.~  Djilkibaev {\it et al.} [Baikal Collaboration], Nucl. Phys. B Proc.Suppl. 143 (2005) 335.

\bibitem{wb}J.~N.~Bahcall and E.~Waxman,Phys.\ Rev.\ D {\bf 64} (2001) 023002; J.~N.~Bahcall and E.~Waxman, Phys.\ Rev.\ D {\bf 59} (1999) 023002.
\bibitem{mpr}K.~Mannheim, R.~J.~Protheroe and J.~P.~Rachen, Phys.\ Rev.\ D {\bf 63} (2001) 023003.

\bibitem{Athar:2000yw}
  H.~Athar, M.~Jezabek and O.~Yasuda,
  Phys.\ Rev.\ D {\bf 62} (2000) 103007.


\bibitem{costa}
  C.~G.~S.~Costa,
  Astropart.\ Phys.\  {\bf 16} (2001) 193.

\bibitem{showerpower}  J.~F.~Beacom and J.~Candia, JCAP {\bf 0411}, 009 (2004).

\bibitem{Candia:2003ay}
  J.~Candia and E.~Roulet,
  JCAP {\bf 0309}, 005 (2003).

\bibitem{lipari} P.~Lipari, Astropart.\ Phys.\ \textbf{1} (1993) 195.

\bibitem{hooper}
  D.~Hooper, H.~Nunokawa, O.~L.~G.~Peres and R.~Zukanovich Funchal,
  Phys.\ Rev.\ D {\bf 67} (2003) 013001.


\bibitem{honda} T.~K.~Gaisser and M.~Honda,
  Ann.\ Rev.\ Nucl.\ Part.\ Sci.\  {\bf 52} (2002) 153.

\bibitem{pdg} S.~Eidelman et al. [Particle Data Group], Phys. Lett. B{\bf 592} (2004), 1.


\bibitem{shigeru}
  S.~Yoshida, R.~Ishibashi and H.~Miyamoto,
  Phys.\ Rev.\ D {\bf 69} (2004) 103004.

\bibitem{anis}A.~Gazizov, M.~Kowalski, submitted to Comp.~Phys.~Commun., Preprint arXiv:astro-ph/0406439.


\bibitem{prem} see Ref.~[83] in R.~ Gandhi et al., Astropart.\ Phys.\ \textbf{5}
 (1996) 81.
\bibitem{cteq} H. L.~La  et al. [CTEQ Collaboration], hep-ph/9903282; \\
http://www.phys.psu.edu/$\sim$cteq/

\bibitem{tauola} S.~Jadach et al., Comput.\ Phys.\ Commun.\
\textbf{ 76} (1993) 361.

\bibitem{doublebang}
  J.~G.~Learned and S.~Pakvasa,
  Astropart.\ Phys.\  {\bf 3} (1995) 267.

\bibitem{marek_phd}M. Kowalski, PhD Thesis, Humboldt University, Berlin (2003); \hfill \\ http://area51.berkeley.edu/manuscripts/.

\bibitem{fisher} R.A.~Fisher, J.~Roy.~Stat.~Soc. 98 (1935) 39.  
\bibitem{tegemark} M.~Tegmark, A.~Taylor, A.~Heavens, Astrophys.~J. 480 (1997) 22.



\bibitem{heiko}H. Geenen et al. [AMANDA Collaboration], published in Proc. of ICRC2003, Tsukuba, Japan (2003).
\bibitem{ped}P. Miocinovic, PhD Thesis, University of California, Berkeley (2001);\hfill \\ http://area51.berkeley.edu/manuscripts/.


\bibitem{beacom_tau}
  J.~F.~Beacom, N.~F.~Bell, D.~Hooper, S.~Pakvasa and T.~J.~Weiler,
  Phys.\ Rev.\ D {\bf 68}, (2003) 093005. 




\end{thebibliography}
\end{document}